\documentclass{tcibook}
\usepackage{fancyhea}
\usepackage{sidecap}

\usepackage{work}
\usepackage{bm}       
\usepackage{graphicx}
\usepackage{hyperref}      
\usepackage{chngcntr}
\counterwithout{figure}{chapter}

\newcommand{\nc}{\newcommand}  

\def\beq{\begin{equation}}
\def\eeq#1{\label{#1}\end{equation}}
\def\eeqn{\end{equation}}

\newenvironment{Eqnarray}%
   {\arraycolsep 0.14em\begin{eqnarray}}{\end{eqnarray}}
\def\beqa{\begin{Eqnarray}}
\def\eeqa#1{\label{#1}\end{Eqnarray}}
\def\eeqan{\end{Eqnarray}}

\nc{\ra}{\rightarrow}  
\nc{\slsh}{\slash\hspace*{-0.22cm}}
\def\Re{{\cal R \mskip-4mu \lower.1ex \hbox{\it e}\,}}
\def\Im{{\cal I \mskip-5mu \lower.1ex \hbox{\it m}\,}}

\nc{\vev}[1]{ \left\langle {#1} \right\rangle }
\nc{\bra}[1]{ \langle {#1} | }
\nc{\ket}[1]{ | {#1} \rangle }
\nc{\fb}{\,{\rm fb}^{-1}}
\nc{\ev}{{\rm eV}}
\nc{\kev}{{\rm keV}}
\nc{\Mev}{{\rm MeV}}
\nc{\gev}{{\rm GeV}}
\nc{\tev}{{\rm TeV}}
\nc{\mev}{{\rm MeV}}

\def\del{\partial}
\def\Dslash{\not{\hbox{\kern-4pt $D$}}}
\def\dslash{\not{\hbox{\kern-2pt $\del$}}}
\def\pslash{\not{\hbox{\kern-2pt $p$}}}
\def\ETmiss{ \not{\hbox{\kern-4pt $E$}}_T }

\def\msb{{\bar{\ssstyle M \kern -1pt S}}}

\begin{document}

\def\bibname{References}

\bibliographystyle{utphys}  

\raggedbottom

\pagenumbering{roman}

\parindent=0pt
\parskip=8pt
\setlength{\evensidemargin}{0pt}
\setlength{\oddsidemargin}{0pt}
\setlength{\marginparsep}{0.0in}
\setlength{\marginparwidth}{0.0in}
\marginparpush=0pt


\pagenumbering{arabic}

\renewcommand{\arraystretch}{1.25}
\addtolength{\arraycolsep}{-3pt}


\newcommand{\sfig}[2]{
\centering
\includegraphics[width=#2]{#1}
}
\newcommand{\Sfig}[2]{
    \begin{figure}[htbp]
    \sfig{#1.pdf}{0.55\columnwidth}
   \caption{#2} 
    \label{fig:#1}
    \end{figure}
}
\newcommand{\rf}[1]{~\ref{fig:#1}}
 
\chapter*{Dark Energy and CMB}

\begin{center}\begin{boldmath}



\begin{center}

\begin{large} {\bf Conveners: S.~Dodelson and K.~Honscheid} \end{large}

\begin{large} {\bf Topical Conveners:} \end{large} K.~Abazajian,
J.~Carlstrom,
D.~Huterer,
B.~Jain,
A.~Kim,
D.~Kirkby,
A.~Lee,
N.~Padmanabhan,
J.~Rhodes,
D.~Weinberg

\end{center}



\end{boldmath}\end{center}

\section*{Abstract}

The American Physical Society's Division of Particles and Fields initiated a long-term planning exercise over 2012-13, with the goal of developing the community's long term aspirations. The sub-group ``Dark Energy and CMB'' prepared a series of papers explaining and highlighting the physics that will be studied with large galaxy surveys and cosmic microwave background experiments. This paper summarizes the findings of the other papers, all of which have been submitted jointly to the arXiv.


\newpage
\renewcommand*\thesection{\arabic{section}}
\section{Cosmology and New Physics} 

Maps of the Universe when it was 400,000 years old from observations of the cosmic microwave background and over the last ten billion years  from galaxy surveys point to a compelling cosmological model. This model requires a very early epoch of accelerated expansion, inflation, during which the seeds of structure were planted via quantum mechanical fluctuations. These seeds began to grow via gravitational instability during the epoch in which dark matter dominated the energy density of the universe, transforming small perturbations laid down during inflation into nonlinear structures such as million light-year sized clusters, galaxies, stars, planets, and people. Over the past few billion years, we have entered a new phase, during which the expansion of the Universe is accelerating presumably driven by yet another substance, dark energy.

Cosmologists have historically turned to fundamental physics to understand the early Universe, successfully explaining phenomena as diverse as the formation of the light elements, the process of electron-positron annihilation, and the production of cosmic neutrinos. However, the Standard Model of particle physics has no obvious candidates for inflation, dark matter, and dark energy. The amplitude of the perturbations suggest that the natural scale for inflation is at ultra-high energies\footnote{Roughly, the measured amplitude of the density perturbations $\delta\rho/\rho \simeq 10^{-5} \sim (E_{\rm inf}/m_{\rm Planck})^2/\sqrt{\epsilon}$, where $\epsilon\simeq 0.01$ is a small parameter in slow roll inflation.}, so understanding the physics driving inflation could lead to information about the UV completions of our current theories. There are arguments that naturally link the dominant dark matter component to new physics hovering above the electroweak scale, and the powerful suite of experiments aiming to find this component(s) were the focus of several separate groups in the Snowmass process. Apart from the dominant component, neutrino oscillation experiments already inform us that neutrinos constitute a non-negligible fraction of the dark matter, and an important message is that experiments usually associated with dark energy and inflation are ideally suited to pin down the sum of the masses of the neutrinos and the cosmic existence of any additional (sterile) species. The situation with dark energy is more complex. 
A cosmological constant ($\Lambda$) has effective pressure equal to minus its energy density (equation of state $w=-1$) consistent with preliminary measurements, but for example in supersymmetric theories the most natural scale for $\Lambda$ is at least as large as 100 GeV. A cosmological constant with this value would produce a universe accelerating so rapidly that the tips of our noses would be expanding away from our faces at a tenth the speed of light. If $\Lambda$ is responsible for the current epoch of acceleration, its value is many orders of magnitude smaller than this but curiously just large enough that it began dominating the energy density of the universe only recently. So the mechanism driving the current accelerated expansion of the Universe remains a profound mystery.

The quest to understand dark energy, dark matter, and inflation then is driven by a fundamental tension between the extraordinary success of the model that explains our Universe and the failure of the Standard Model of particle physics to provide suitable candidates for the dark sector that is so essential to our current view of the Universe. Experiments on the cosmic frontier have demonstrated that the Standard Model is incomplete; the next generation of experiments can provide the clues that will help identify the new physics required. 
Complementing the efforts on the Intensity and Energy Frontier, physicists working on the Cosmic Frontier are poised to unravel the mysteries of the cosmos.

\section{Dark Energy}

Physicists have proposed a number of different mechanisms that could be responsible for the accelerated expansion of the Universe.
None is compelling, but some of them have been predictive enough to fail, while others have led to a deeper understanding of field theory and gravity. Apart from the cosmological constant solution itself, all are predicated on the assumption that there is some (unknown) mechanism that sets $\Lambda$ to zero. 

One possibility is that there is a previously undiscovered substance that contributes to the energy density of the Universe in such a way that the expansion accelerates. In the context of general relativity, acceleration (the positive second derivative of the scale factor $a$) is governed by Einstein's Equations, which reduce to
\begin{equation}
\frac{\ddot a}{a} = - \frac{4\pi G}{3}\,\left[ \rho + 3P\right]
\end{equation}
where $G$ is Newton's constant, $\rho$ the energy density, and $P$ the pressure.
The substance that drives acceleration, therefore, must have negative pressure, or equation of state $w\equiv P/\rho < -1/3$. A nearly homogeneous scalar field whose potential energy dominates over its kinetic energy satisfies this requirement, so many {\it quintessence} models emerged with potentials designed to fit the data. In almost all viable models, however, the mass of the field must be less than the Hubble scale today, of order $10^{-33}$ eV. Embedding such a field in some extension of the Standard Model therefore is challenging as one would expect the scalar mass to get loop corrections many orders of magnitude larger than this. This mass stability problem is an indication of just how hard the problem is: the mass that is protected is some 44 orders of magnitude smaller than the Higgs mass that lies at the center of the electroweak hierarchy problem. It seems clear that the new physics is an infrared phenomenon as opposed to all prior hints of new physics, which have entered from the ultraviolet.

With no real guidance from theory, quintessence models are nevertheless appealing because they open up the parameter space: most models have $w\ne -1$, and many have evolving equations of state so that $dw/da$ is also non-zero. This class of models therefore offers a clear, and arguably more appealing, alternative to the cosmological constant that would be favored if surveys find deviation from $w=-1$.
Although the quintessence field does not clump on scales smaller than its (very large) Compton wavelength, on the largest scales dark energy should clump, yet another difference from the cosmological constant model. Finally, some models allow for episodic dark energy domination, so that the present accelerating era is not particularly special. Indeed, an early epoch of inflation is one such epoch, but there may have been others, for example during phase transitions. Measuring the effects of dark energy in a series of redshift\footnote{Redshift $z\equiv a^{-1} - 1$; high redshift corresponds to early cosmic time.} bins is therefore necessary to distinguish among the many possibilities.
This is an area where the cosmic microwave background, whose lensing maps are sensitive to structure from $z=10^3$ until today, can be profitably combined with galaxy surveys, whose lensing maps probe structure at a sequence of lower redshifts. 

Quintessence models explicitly introduce an extra degree of freedom in the form of a new scalar field. Early attempts~\cite{Carroll:2003wy} to modify gravity (the ``left-hand side of Einstein's Equations'') implicitly introduced an extra degree of freedom~\cite{Chiba:2003ir}, but in a slightly different way. Technically, these ``scalar-tensor'' models differ from quintessence in that the coefficient of the Ricci scalar $R$ in the Jordan frame\footnote{Theories can be written either in the Jordan frame, in which matter couples to the metric only through $\sqrt{-g}\mathcal{L}_m$, or in the Einstein frame, in which the gravitational part of the action remains Einstein-Hilbert and there are additional couplings of the new degrees of freedom to matter.} in the action depends on the new field. Different scalar-tensor models can therefore produce the variety of predictions found in quintessence models, but they extend the possibilities in a new direction: the non-canonical coupling to the Ricci scalar propagates to the equations that govern the evolution of perturbations. This leads to the general conclusion that modified gravity models typically predict that structure grows at a different rate (e.g., \cite{Hu:2007pj}) than in models based on general relativity. Differentiating between modified gravity and General Relativity-based dark energy therefore will require both measurements of cosmological distances to pin down the background expansion and then measurements of the growth of structure to distinguish between them. 

Although there have been many proposals for how to modify gravity, the most interesting development recently traces back to an idea first proposed by Fierz and Pauli~\cite{Fierz:1939ix}, that the graviton has non-zero mass. Qualitatively a massive graviton seems like an appropriate way to decrease the strength of the gravitational force on very large scales and hence to explain the acceleration of the Universe. In practice, the theory runs into two problems, both related to the fact that a massive spin-2 particle carries degrees of freedom beyond those of the massless graviton. These extra degrees of freedom typically lead to modifications to general relativity in the Solar System, modifications that are excluded by the tight limits on post-Newtonian parameters. The second challenge for massive graviton models is to avoid Boulware-Deser ghosts~\cite{Boulware:1973my}, the instability of one of the extra degrees of freedom. 

To satisfy the Solar System constraints, the extra degrees of freedom need to be {\it screened}, i.e. heavily suppressed by limiting their range of interaction or effective coupling to matter in environments like the Solar System. Any successful modified gravity model needs a screening mechanism. One possibility is Vainshtein screening~\cite{Vainshtein:1972sx}, which arises due to non-linearities in the Fierz-Pauli potential. In general, these non-linear terms do not avoid the ghost problem, but recently~\cite{Nicolis:2008in} the set of terms in the potential that are safe from ghosts have been identified. These {\it Galileon} models offer a potentially attractive way of addressing the acceleration of the Universe within a consistent framework. Beyond this theoretical breakthrough, the Vainshtein screening intrinsic to this model (and other proposed mechanisms) open up yet another axis of tests: how and where do modified gravity theories transition to normal Newtonian gravity in the Solar System? The full suite of ways to test modified gravity models is still under development.

The modern view of the formation of structure in the Universe has been confronted with a growing array of precise tests, including
the anisotropy spectrum (both temperature and polarization) of the cosmic microwave background; light curves of distant Supernovae; abundances of galaxy clusters; clustering of galaxies, quasars, and Lyman alpha systems; gravitational lensing; and cross-correlations between different pairs of these observations. The basic framework, with inflation, dark matter, and dark energy at its core, has been confirmed repeatedly over the past decade. We need to keep pushing: either the basic picture will break or the agreement will become even more remarkable. Thinking in parameter space, the next decade will enable us to reduce the uncertainty in the equation of state by a factor ten. Historically in physics, precision measurements of this sort have been pursued for these ends: will the simple theory hold up or will it need to be replaced by something more profound? In the case of the cosmic acceleration, where there is no appealing fiducial model, the push for greater precision takes on an even greater importance.

Since scientists in the United States discovered evidence for cosmic
acceleration over a decade ago, the US has been the leader in the
field of dark energy studies.  The 2006 Dark Energy Task Force
report~\cite{Albrecht:2006um} provided a systematic discussion of experimental
approaches to dark energy and identified a sequence of ``Stage III''
and ``Stage IV'' dark energy experiments to build on those then in progress. Stage III surveys were designed to address systematics with the goal of 
statistics-limited constraints from four independent probes. Each of these probes will be developed over the ensuing decade to the point at which Stage IV 
surveys will enable sub-percent level consistency checks for all probes, some of which will have reached their cosmic variance limit (i.e., that we have only a single observable
universe to probe).
This staged categorization was reiterated in the more recent
Community Dark Energy Task Force Report~\cite{rocky3}, which particularly emphasized the importance of
complementing planned imaging experiments with spectroscopic experiments.
Fig.~\rf{Facilities} illustrates the timelines for several of the major
dark energy experiments in which US scientists are playing an important or leading role with further details provided in a separate 
document~\cite{projects}.  

\Sfig{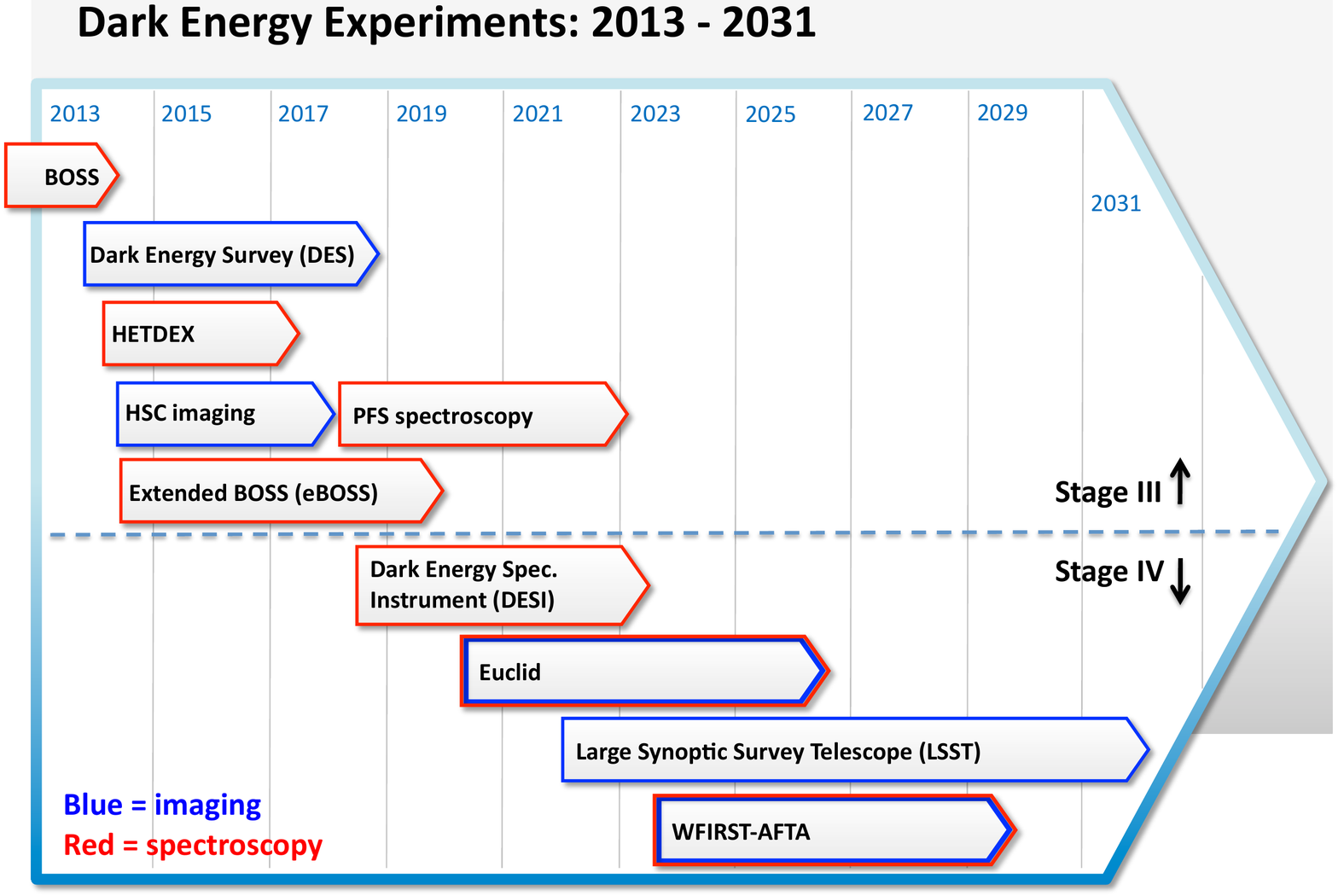}{A timeline of Stage III and Stage IV dark energy experiments -- photometric and spectroscopic -- in which US scientists are
playing an important or leading role.  
Most of the projects are ground-based with either US leadership (BOSS, DES, HETDEX, eBOSS, DESI, LSST) or active participation (HSC, PFS). The two space missions are Euclid, led by the ESA with a
NASA-sponsored team of US participants, and WFIRST, led by NASA.}

Major gains beyond the current road map of dark energy projects will require advances on one of a number of fronts.  One potential avenue is to develop new techniques or probes of cosmology and new physics that have not yet been developed; data from many cosmological surveys have been used for tests not envisioned when those experiments were first designed, and we anticipate that that trend will continue ~\cite{novel}.  Another path is to obtain complementary information that will enhance planned experiments; for instance, a targeted program of spectroscopic redshift measurements can enhance the dark energy constraints from LSST compared to its baseline capabilities~\cite{Zhan:2006gi}.  Finally, new instrumental capabilities -- e.g., methods that suppress flux from emission lines from the night sky, which are much brighter than distant galaxies at infrared wavelengths, or new detector technologies that would measure photon energy rather than just quantity -- could enable powerful new dark energy experiments to be conducted at reduced costs.  Modest investments in these avenues may yield large potential payoffs in the future.

The following sub-sections summarize the physical probes and measurements
that speak to dark energy science.  Each of these approaches is detailed
further in a separate document~\cite{distance,growth,novel,cross}.  Taken together, these measurements
and the projects that enable them constitute an all-out attack on the
problem of cosmic acceleration.  The program planned over the coming
decade guarantees continuing US leadership in this field.


\subsection{Distances}

The relationship between the redshift and distance of an object is one of the primary tests of the expansion history of the Universe, and therefore played a key role in the discovery of the accelerating Universe. The simple graph of the distance scale of the Universe as a function of redshift, indicating the evidence for cosmic acceleration, has become an iconic plot in the physical sciences.
The data for this plot so far come from measurements of Type Ia supernovae and baryon acoustic oscillations (BAO)
and these will be the sources of streams of data in coming years. DES and LSST will provide an essentially
limitless supply of supernova, thousands, then hundreds of thousands.
The DES collaboration and LSST-DESC will coordinate the spectroscopic classification of a
fraction of these objects.
The challenge is to make measurements
thoroughly enough to mitigate systematic problems, especially those that depend on redshift. Detailed
studies of nearby supernovae are beginning to provide clues for how to do this. Much would be gained if
observations could be made from space, but a substantial gain will be also achieved if we make ground-based
observations that avoid the atmospheric lines in the near infrared.

The subtle pattern of anisotropy in the cosmic microwave background, just one part in $10^5$, is mapped at the two
dimensional boundary of a three-dimensional feature, the fluctuations in matter density throughout space.
The counterpart of the oscillations in the CMB power spectrum is a peak in the correlation between the
densities at points separated by 153 Mpc measured relative to the scale size of the Universe, left behind by
baryon acoustic oscillations in the early Universe. This very large meter stick can be observed as far out as redshifts
$z = 1.6$ using galaxies as traces of matter density, and even out to $z = 3$ using light from
quasars. The current Baryon Oscillation Spectroscopic Survey (BOSS) is likely to
report a distance measurement soon with 1\% accuracy. 
The eBOSS survey is designed to extend the reach of BAO measurements to higher redshifts. 
The Stage-IV BAO experiment, DESI, should provide more than 30 similarly accurate independent distance measurements.

\Sfig{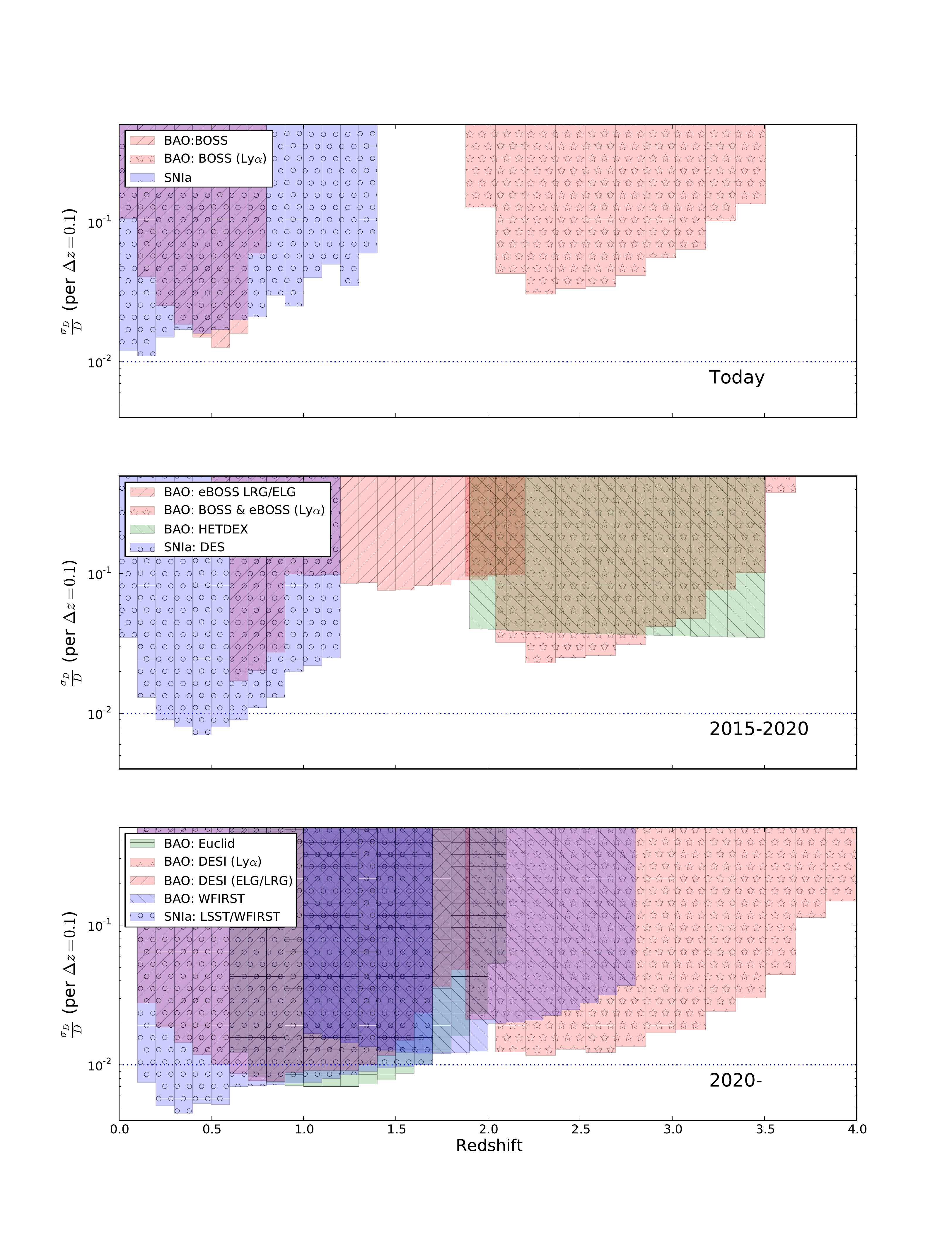}{Current and projection for future uncertainties on cosmic distance as a function of redshift.}

Fig.~\rf{distance} shows the current and projected future constraints on cosmic distances using these two techniques.
If our basic understanding is correct, the supernova and BAO measurements should be in absolute agreement.
The distance-versus-redshift curve of the Universe is fundamental and exploring it with completely different
techniques is essential. These stunning measurements will allow for percent level determination of the equation of state and be extremely sensitive to evolution of the dark energy at earlier times. By pinning down the distance-redshift relation, they will also allow for apples-to-apples comparisons of modified gravity vs. dark energy models using the growth of structure.

\subsection{Growth of Structure}
\def\hmpcinv{\,h\,{\rm Mpc^{-1}}}
\def\hinvmpc{\,h^{-1}{\rm Mpc}}

The quantity and quality of cosmic structure observations have greatly
accelerated in recent years, and further leaps forward will be
facilitated by imminent projects.  These will enable us to map the
evolution of dark and baryonic matter density fluctuations over cosmic
history.  The way that these fluctuations vary over space and time is
sensitive to several pieces of fundamental physics: the primordial
perturbations generated by GUT-scale physics; neutrino masses and
interactions; the nature of dark matter; and dark energy.  We focus on
the last of these here: the ways that combining probes of growth with
those of cosmic distances will
pin down the mechanism driving the acceleration of the Universe.

If the acceleration is driven by dark energy, then 
distance measurements provide one set of constraints on $w$,
but dark energy also affects how rapidly structure grows.
Upcoming surveys are therefore designed to probe $w$ in two distinct ways: direct 
observations of the distance scale and the growth of structure,
each complementing the other on both systematic errors and dark energy constraints.
A consistent set of results will greatly increase the reliability of the final answer. 

\Sfig{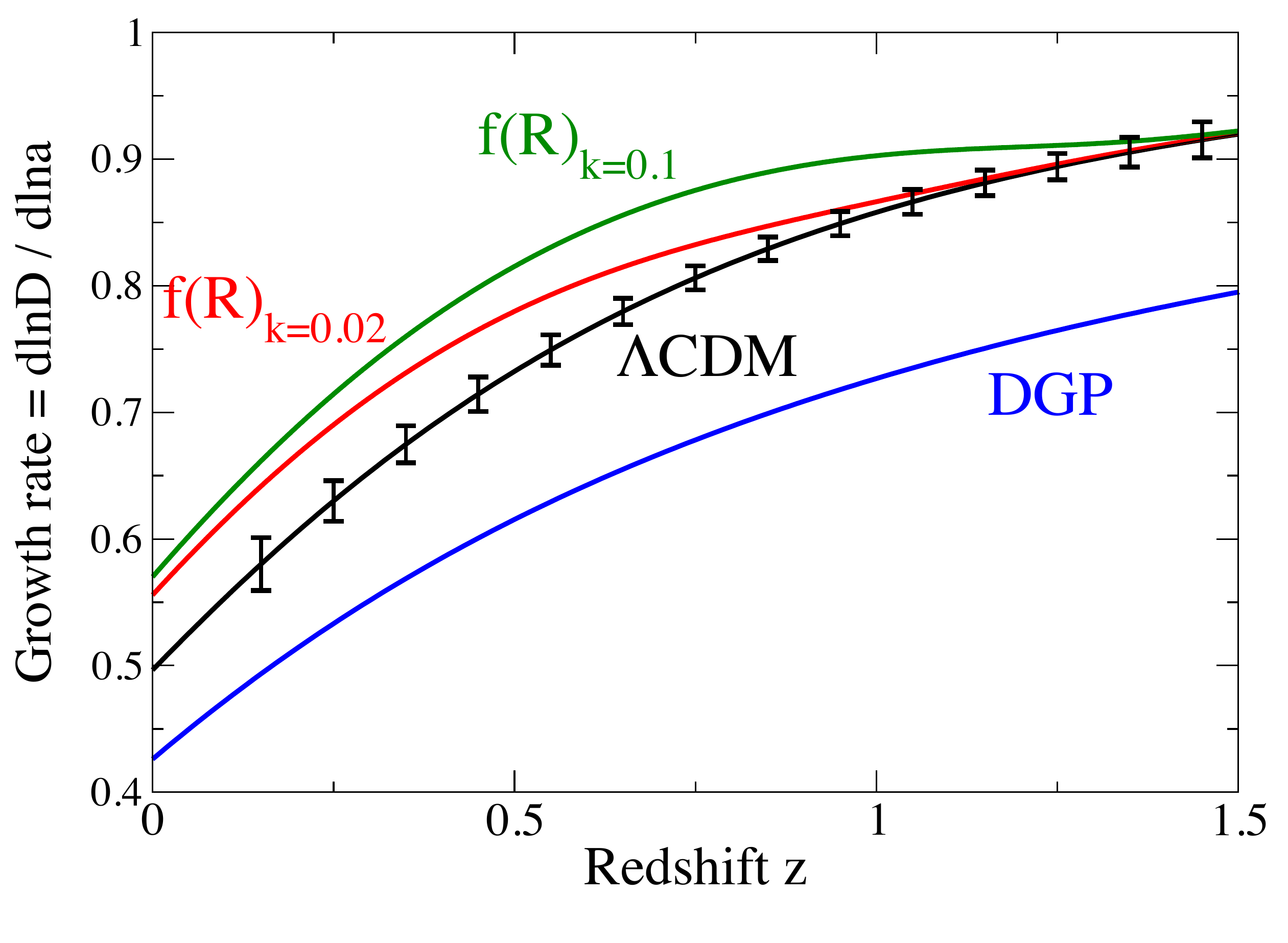}{Constraints on the growth of density fluctuations in the universe 
  with errors projected from DESI. The curves show the derivative of the logarithmic
  growth with respect to logarithmic scale factor --- a quantity readily measured from the
  clustering of galaxies in redshift space --- as a function of redshift. We show theory
  predictions for the standard $\Lambda$CDM model, as well as for two modified-gravity
  models, the Dvali-Gabadadze-Porratti (DGP) model \cite{DGP}, and for
  the $f(R)$ modification to Einstein action \cite{0905.2962}. Because growth
  in the $f(R)$ models is generically scale-dependent, we show
  predictions at wave numbers, $k=0.02\hmpcinv$ and $k=0.1\hmpcinv$. LSST projects to impose constraints of similar excellent quality on the growth function $D(a)$.}

If cosmic acceleration is driven by modified gravity, then probes of structure become even more important.
Generically, modified gravity models are able to reproduce any expansion history that can be attained in dark energy models,
but at the cost of altering the growth of
structure. How rapidly structure grows is quantified by the dimensionless {\it growth function} $D(a)$. Figure \rf{growth_killer} illustrates how different models make predictions that differ from those of general relativity even though the distance-redshift relation in all these models is identical. The growth of structure then will be able to distinguish modified gravity from dark energy as an explanation for the cosmic acceleration. 

Fig.~\rf{growth_killer} projects constraints from a spectroscopic survey that measures the local velocities of galaxies by observing redshift space distortions.  Similarly powerful constraints are projected from photometric surveys that are dedicated to measuring the shapes of galaxies and therefore are sensitive to the signal from weak gravitational lensing.

 Achieving these powerful constraints will require both wide field imaging and spectroscopic
redshift surveys, as depicted in Fig.~\rf{Facilities}.
The results will pin down far more than the equation of state to percent level accuracy, although this in itself will be an important clue as to whether the cosmological constant or an alternative is driving the acceleration of the Universe. We will learn also whether the equation of state is varying with time and whether dark energy was relevant at high redshift. The surveys will probe cosmological perturbations as a function of both length and time, opening up dozens of possible failure modes for GR-based dark energy.  If any of these is reliably detected to differ from the GR prediction, we will have a revolution on our hands.

\subsection{Novel Probes}

Surveys enabling the twin probes of distances and structure can distinguish between modified gravity and dark energy on cosmological scales. It has become apparent over the last few years that non-cosmological tests can also play an important role in determining the mechanism driving the acceleration of the Universe. The basic idea is that gravity is known to reduce to general relativity (GR) on Solar System scales, so any modified gravity model must have a screening mechanism, wherein the additional forces operative on large scales are suppressed in the solar system. Indeed, many of the models have screening built into them, so the solar system constraints can be naturally satisfied. The key issue is the nature of the transition from large (modified gravity) to small (GR) scales, and how that transition can be detected observationally.

Screening mechanisms typically utilize some measure of the mass distribution of halos, such as the density or the Newtonian potential, to recover general relativity (GR) well within the Milky Way. This leaves open the possibility that smaller halos, the outer parts of halos, or some components of the mass distribution, are unscreened and therefore experience enhanced forces. For a given mass distribution, unscreened halos will then have internal velocities and center of mass velocity larger than predicted by GR. Deviations from GR are typically at the ten percent level, with distinct variations between different mechanisms in the size of the effect and the way the transition to GR occurs. It is important to note that observable effects are typically larger on galaxy scales than on large (cosmological) scales or at high redshift.  The comparison of dynamical and lensing masses provides a powerful test that is being implemented on a wide range of scales: from individual galaxies to large-scale cross-correlations that are also discussed in the Growth of Structure and Cross-Correlations sections.

\Sfig{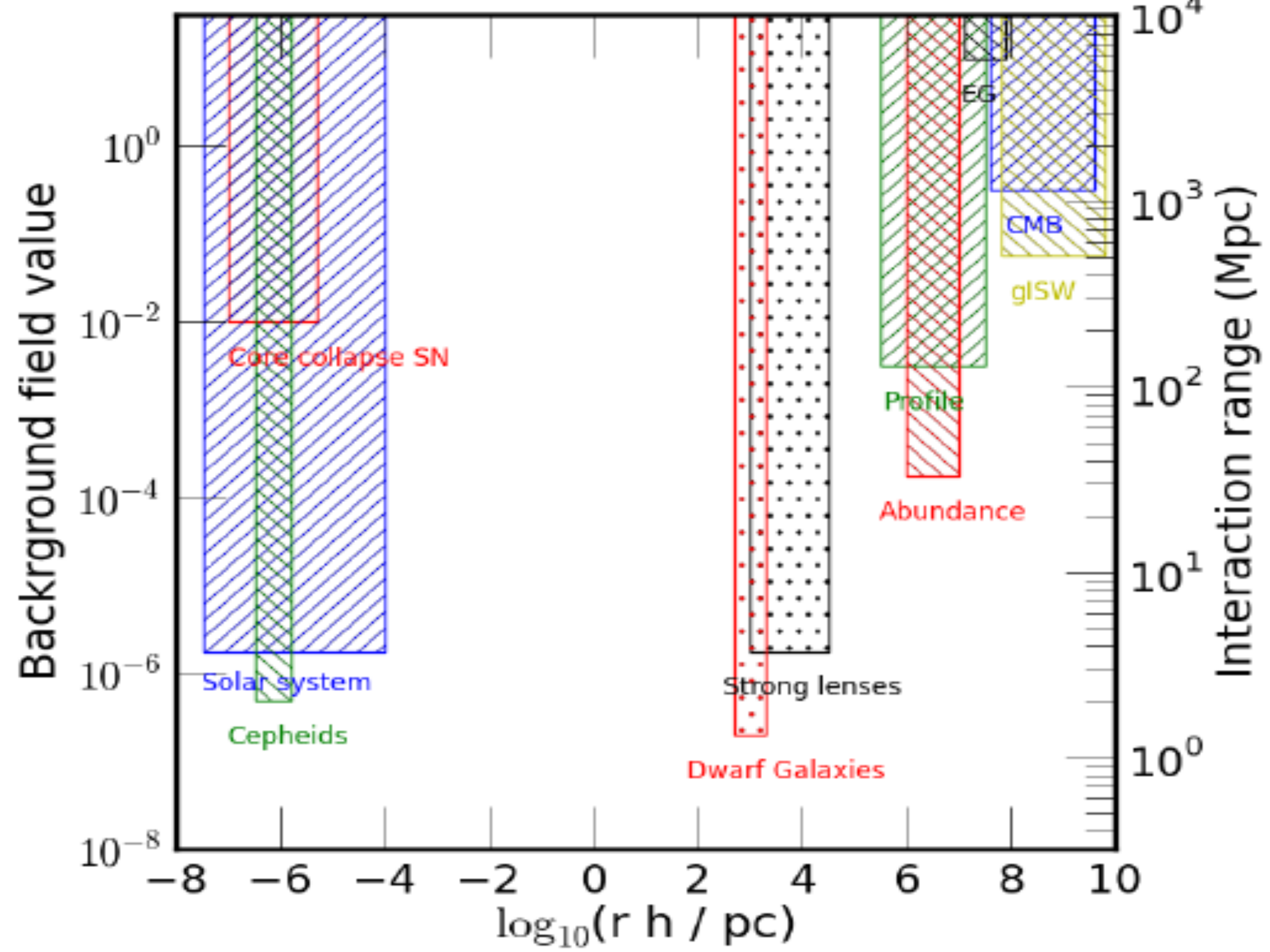}{Astrophysical~\cite{Hu:2007nk,Jain:2012tn,Vikram:2013uba} and cosmological~\cite{Song:2007da,Giannantonio:2009gi,Schmidt:2009am} limits on chameleon theories, in particular $f(R)$ models. The spatial scale on the x-axis gives the range of length scales probed by particular experiments. The parameter on the y-axis is the value of the field that mediates the additional force in units of the Planck mass, or equivalently the range of the additional force. The rectangles show excluded regions; the two rectangles with dots are meant to indicate preliminary results from ongoing work.}

Fig.~\rf{novel} shows a range of tests that have been implemented in one particular model, f(R) gravity, with Chameleon screening. The screening in this model, as in many others, depends on the value of a field in the region of interest. The y-axis in Fig.~\rf{novel} depicts the value of this field in Planck units divided by a coupling constant, while the x-axis shows that the model has been probed on scales ranging from the Solar System all the way out to cosmological scales, with all tests to date verifying GR. In the next decade essentially the entire accessible parameter space of Chameleon theories can be probed using the tests shown in the figure. Tests for the other important screening mechanism, Vainshtein screening, are at early stages, with potential for rapid progress. 

The program of testing gravity theories via these novel probes is in its infancy, but it is becoming increasingly clear that even modest investments in non-cosmological observations have enormous potential to contribute to the cosmic acceleration problem.

\subsection{Cross-Correlations}

To extract the most information possible about dark energy from galaxy surveys and CMB experiments, scientists will be forced to combine probes.
The multi-probe approach is not simply a matter of multiplying uncorrelated likelihoods. As but one example, gravitational lensing and large scale structure are highly correlated probes, with the lensing signal enhanced behind over-dense regions and the density enhanced (due to magnification) behind regions with large lensing signals. So the likelihoods are not independent, and there is significant information contained in the cross-correlations. Beyond improvement in statistical power, these cross-correlations will enable the community to isolate systematics. 
The four principal probes (supernovae, clusters, lensing, and large scale structure) are correlated with one another in ways that will need to be accounted for as we strive for percent level accuracy in dark energy parameters.

The four probes enabled by optical surveys will be supplemented by data at other wavelengths. One of the most promising for the purposes of dark energy will come from the oldest probe, the cosmic microwave background. To date, CMB experiments have supplemented galaxy surveys mainly by constraining parameters that would otherwise be degenerate with those that characterize dark energy. Indeed, it is often said that -- since it is sensitive to early universe physics -- the CMB probes dark energy only through the geometric projection from the surface of last scattering.
The photons from the last scattering surface, however, experience deflections due to gravitational lensing and to Compton scattering off free electrons. These deflections show up as {\it secondary anisotropies}, which often do carry information about dark energy.

\Sfig{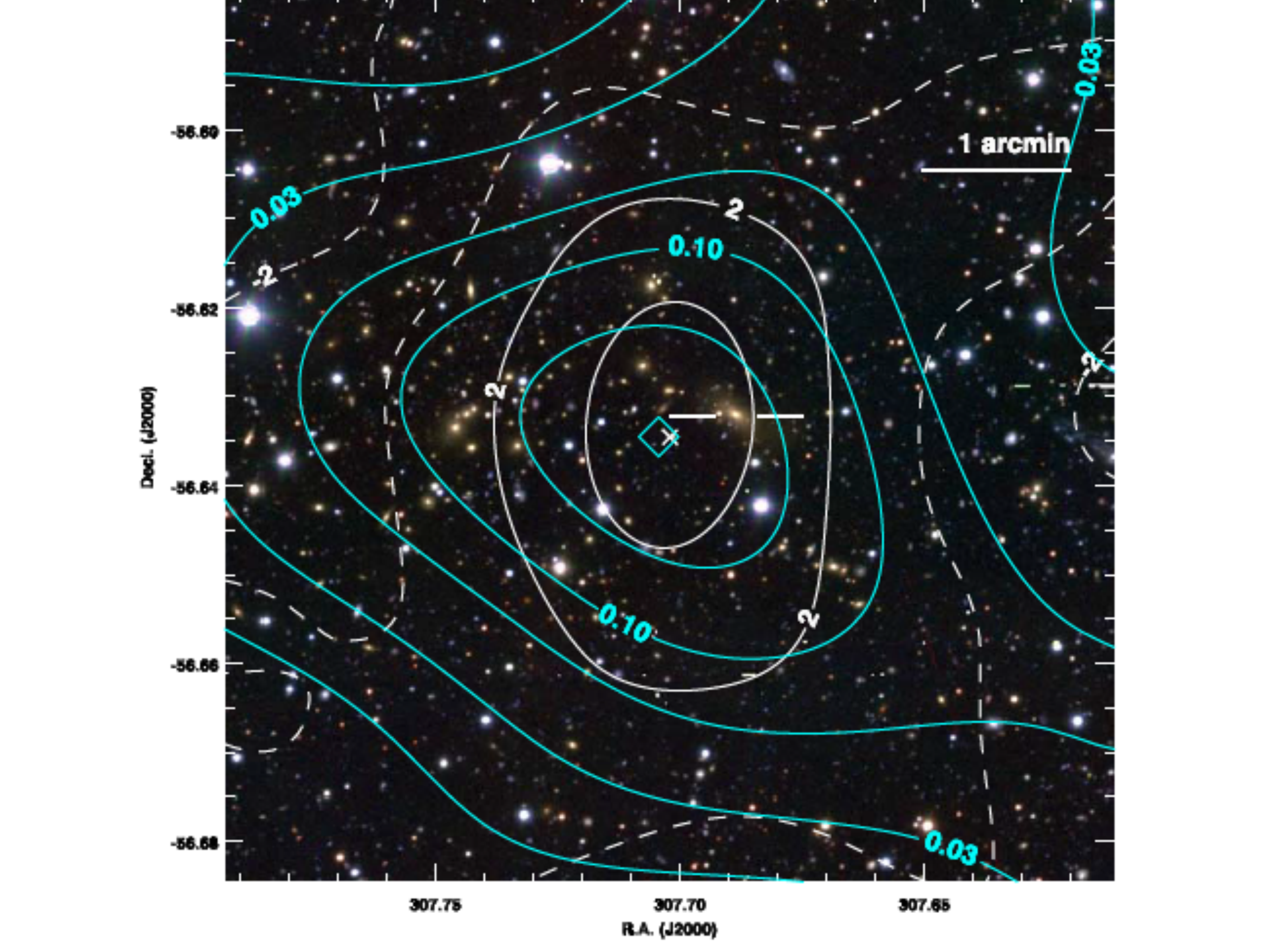}{Map of a galaxy cluster~\cite{High:2012un} using three probes: (i) weak gravitational lensing (blue contours with labels showing the projected density $\kappa$); (ii) hot gas as measured by the Sunyaev-Zel'dovich distortion of the CMB (white contours with labels giving signal to noise); and (iii) galaxies as observed in three optical bands (background).}


Galaxy clusters are perhaps the quintessential example of the value of observing cosmic phenomena at different wavelengths. One of the key uses for clusters in dark energy studies is to compare the abundance above a given mass threshold as a function of redshift with theoretical predictions. The key uncertainty remains the mass determination, and it is in this regard that multi-wavelength studies become particularly important. By observing clusters in optical surveys, in the microwave via the Sunyaev-Zel'dovich effect that follows from Compton scattering off hot electrons in the clusters, and in the X-ray when those same electrons emit radiation, the community obtains many possible mass proxies; taken together they offer a powerful attack on the dominant systematic in dark energy cluster studies. Fig.~\rf{sptcl} shows an example of these different views of a galaxy cluster.


\section{Inflation}\label{inflation}

Cosmic inflation is the leading theory for the earliest history of the universe and for the origin of structure in the universe.
Current observations of the large-scale distributions of dark matter and galaxies in the universe and measurements of the Cosmic Microwave Background (CMB) are in stunning agreement with the predictions of inflation. 
The next generations of experiments in observational cosmology are poised to explore the detailed phenomenology of the earliest moments of the universe. 

\Sfig{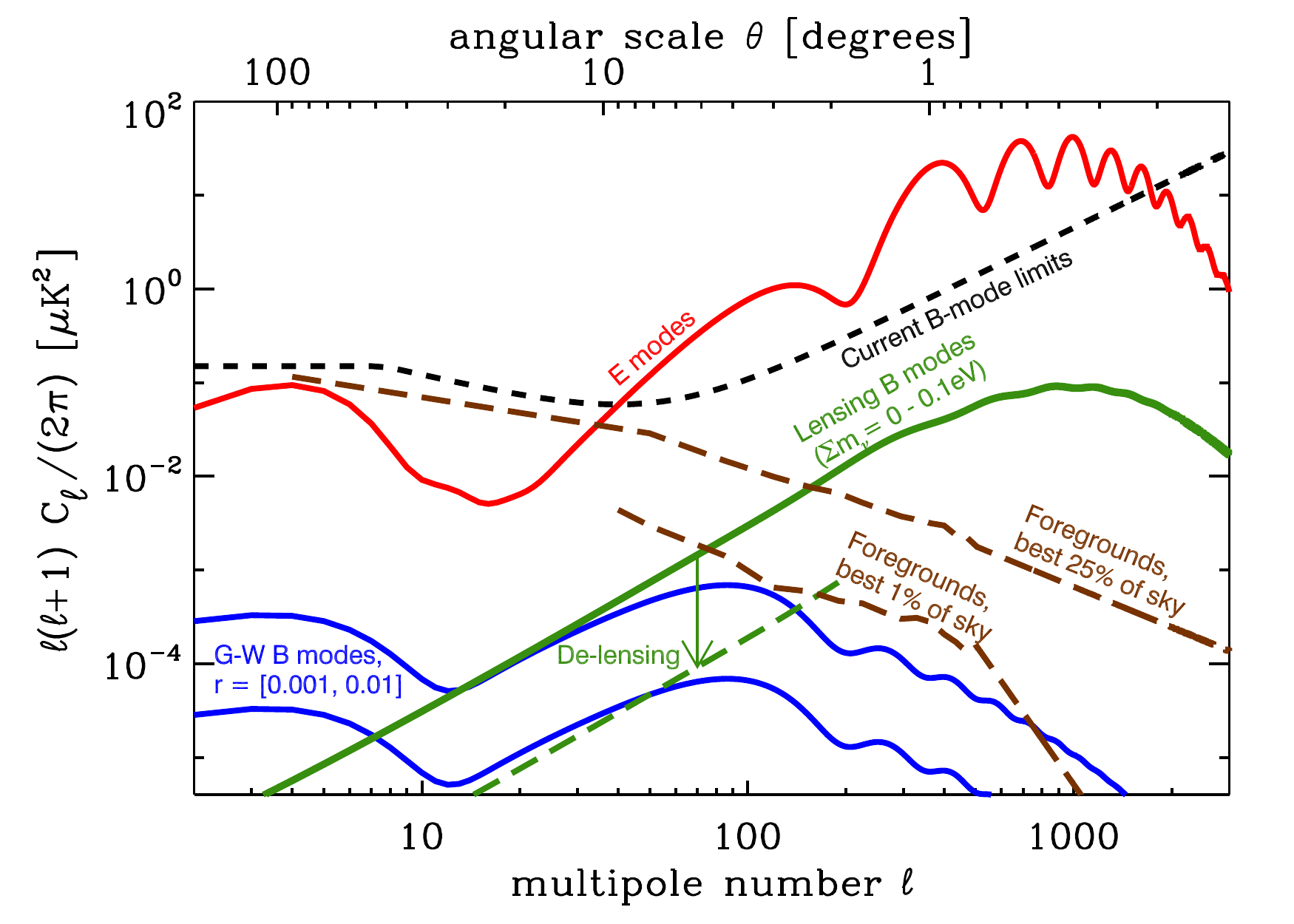}{Expected signal levels for the CMB Polarization $E$-mode (\textbf{red, solid}),
inflationary gravity-wave $B$-mode (\textbf{blue, solid}),
and lensing $B$-mode  (\textbf{green, solid}) signals.
The gravitational wave $B$-mode signals are shown for tensor-to-scalar
ratios of $r=0.001$ (the Stage IV goal) and $r=0.01$ (the boundary between
small-field and large-field inflation models). The
lensing $B$-mode signal is shown as a band encompassing the predicted signal for
values of the sum of neutrino masses $0 \le \sum m_\nu \le 0.1 \mathrm{eV}$.
De-lensing by a factor of 4 in amplitude is shown schematically by the \textbf{green arrow},
with the residual signal at $\ell \le 200$ (where the de-lensing is critical to the constraint
on $r$) shown by the \textbf{green, long-dashed} line.
The \textbf{black, short-dashed} line shows the level of current 95\% upper
limits on $B$ modes from \textsc{WMAP}, \textsc{BICEP}, \textsc{QUIET} and \textsc{QUaD} experiments. The
\textbf{brown, long-dashed} lines show the expected
polarized foreground contamination at 95~GHz for the cleanest $1\%$ and $25\%$ of the sky.}

While the landscape of possible models for inflation is large, the theoretical underpinnings
are well understood, and we are able to make concrete predictions for observable quantities.
One key prediction is the existence of a background of gravitational waves from inflation that 
produce a distinct signature in the polarization of the CMB (the $B$-modes in Fig.~\rf{cmb_powspec_for_snowmass}). Under the 
so-called Lyth condition~\cite{Lyth:1996im}, all models in which the field driving inflation varies by an amount of order $m_{\rm Planck}$ will produce gravitational wave
(tensor) fluctuations that are at least $1 \%$ of the amplitude of density-fluctuation (scalar)
power in the CMB ($r=0.01$ in the figure). 
Definitive evidence one way or another as to the presence of tensor modes with amplitudes at or above this level
would therefore serve as a lever to an infinite
sequence of Planck suppressed operators.  Such
scales require a quantum gravity treatment and this will test string-theoretic mechanisms
for large field inflation.  If not detected
it at least decides between two broad classes of models, since it reaches the Planck scale threshold.
This motivates the design of a next-generation CMB experiment with
the sensitivity and systematics control to detect such a signal with at least $5 \sigma$ 
significance, thus ensuring either a detection of inflationary gravitational waves or the ability to 
rule out large classes of inflation models~\cite{inflation}. 


There are several other handles on the physics of inflation. 
Inflation generically predicts small deviations from a scale-invariant spectrum, and current measurements confirm this prediction at the 5-sigma level. 
DESI projects to obtain a 15-sigma detection thereby further reducing the range of allowed models. BOSS, eBOSS, and DESI will
potentially constrain the {\it running} of the primordial spectrum (deviation from a pure power law) at the 0.2\% level, a factor of
five tighter than current constraints~\cite{Ade:2013uln}.



Non-gaussianity of the primordial perturbations can take many forms. The search for one -- so-called {\it local} non-gaussianity --  is particularly important because single field models of inflation generically predict negligible local non-gaussianity~\cite{inflation}, so any detection will falsify a large class of models. Planck has placed strong upper limits on this and other forms of non-gaussianity consistent with these predictions.
The upcoming surveys eBOSS, DESI, and LSST will constrain a variety of forms of primordial non-gaussianty on different spatial scales and be subject to different systematics than the CMB. They will therefore pave the way for even more stringent bounds on inflationary models.



\section{Neutrinos}\label{neutrinos}

One of the most remarkable aspects of physical cosmology is that the study of the largest physical structures in the Universe can reveal the properties of particles with the smallest known cross-section, the neutrinos. 
At the simplest level, this cosmological sensitivity to neutrino properties is due to the fact that
the neutrino cosmological number density is so large as to be second only to CMB photons.
More specifically, the properties of neutrinos alter the effective energy density of cosmological radiation
and therefore the amplitude, shape and evolution of matter
perturbations, leading to changes in observables in the CMB anisotropies and in measures of large-scale structure.

\newcommand\neff{N_{\rm eff}}

The CMB and large scale structure (LSS) measured in galaxy surveys are sensitive to the sum of the neutrino masses and the number of species produced in the early Universe (dubbed $\neff$). These observations are therefore complementary to laboratory probes of neutrinos, which measure mass differences and potentially CP-violation. CMB experiments are sensitive enough to the neutrino energy density to rule out $\neff=0$ at more than 10-sigma; that is, these experiments have already (indirectly) detected the cosmic neutrinos. Together with priors on the redshift-distance relation from galaxy surveys like BOSS, these experiments also place a stringent upper limit on the sum of the neutrino masses, currently around 0.23 eV~\cite{Ade:2013zuv}.

A global fit to solar and atmospheric neutrino flavor oscillations in the standard 3-generation model
determines two mass differences so a third parameter, which can be taken as either the sum of the masses or the lightest mass, is unknown. Due to a sign ambiguity in one of the mass differences, there are two discrete possibilities (normal and inverted hierarchy as depicted in Fig.~\rf{numass_combine2}) for the relationship between these two parameters.
As indicated in Fig.~\rf{numass_combine2}
upcoming CMB experiments and LSS surveys will unambiguously detect the sum of the masses if the hierarchy is inverted and will likely do so at greater than 3-sigma even if nature has chosen the normal hierarchy. 
Measures of the power spectrum from DESI with a Planck prior could improve the current constraint on  $\sum m_\nu$  to 17 meV, and a Stage IV CMB survey combined with BAO measurement from DESI project to achieve similar sensitivity~\cite{neutrino}.
%
%
Fig.~\rf{numass_combine2} highlights another aspect of the complementarity of laboratory experiments and cosmological probes: if the sum of the masses is found to be $0.15$ eV by the latter, then it will take the former to determine the lightest neutrino mass by identifying whether the hierarchy is normal or inverted.

\Sfig{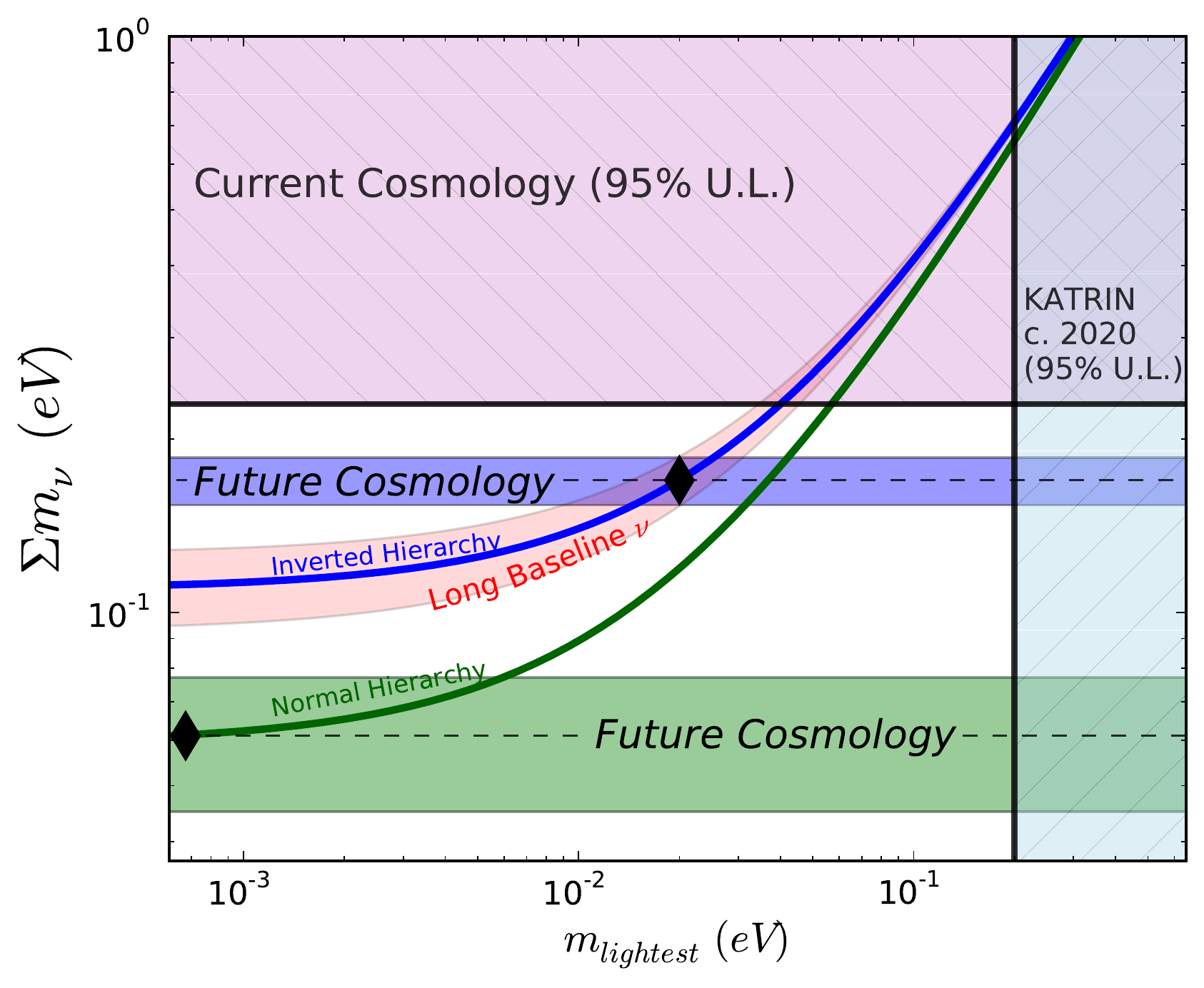}{Current constraints and forecast sensitivity of cosmology to the sum of neutrino masses. In the case of an ``inverted hierarchy,'' with an example case marked as a diamond in the upper curve, future combined cosmological constraints would have a very high-significance detection, with 1-$\sigma$ error shown as a blue band.
In the case of a normal neutrino mass hierarchy with an example case marked as diamond on the lower curve, future cosmology would still detect the lowest $\sum m_\nu$ at greater than 3-$\sigma$.}

Short baseline neutrino oscillation results hint at a richer neutrino sector than three active neutrinos participating in flavor oscillations, with one or more sterile flavors also participating~\cite{Aguilar:2001ty,AguilarArevalo:2010wv,Kopp:2011qd,Giunti:2011gz}. 
These same future CMB experiments will achieve a 1-sigma error of $\Delta\neff=0.027$, which will complement future sterile neutrino searches and inform model building (since sterile neutrinos might be detected in the laboratory even though they were not produced in the early universe and vise versa).

\section{Conclusions}

Cosmological surveys are sensitive to fundamental physics. To date, this basic fact has led to the discovery of the accelerating universe, strong evidence for an epoch of early acceleration near the GUT scale, the indirect detection of the cosmic neutrino background, and the most compelling evidence for non-baryonic dark matter. However, surveys to date have measured only a fraction of all information available. If the Universe were contained in an area the size of the United States, galaxy maps so far would have surveyed the city of Birmingham, Alabama. CMB experiments have provided low-noise maps of the temperature down to angular scales of order a tenth of a degree. Strategic, valuable information remains unmined in higher resolution temperature maps and the virtually uncharted polarization field. We have outlined the projections for how this extra information will constrain dark energy, neutrinos, and inflation; these projections are extraordinary. But even they ignore the very real possibility that future experiments on the cosmic frontier will do just what their predecessors have done: discover something fundamentally new!

The community has rallied behind previous reports~\cite{Albrecht:2006um,Albrecht:2009ct,rocky3} which are consistent with the current consensus to support the following key steps:

\begin{itemize}

\item REMAIN A LEADER IN DARK ENERGY RESEARCH

The U.S. played the leading role in discovering the acceleration of the Universe, as was recognized by the 2011 Nobel 
Prize. The acceleration remains a mystery, whose solution may usher in a revolution in either our theory of gravity 
or our understanding of particle physics. Different classes of theories make different predictions for the growth of 
structure given a redshift-distance relation. A combination of spectroscopic and photometric surveys can determine both
distances and structure growth, so will help pinpoint the new physics driving the acceleration of the universe. The current
suite of surveys, Stage III, will be the first to implement the vision of multiple probes and small systematics. 
This vision will be realized fully with the Stage IV surveys (LSST and
DESI), as they reach the level where exquisite-precision dark energy
constraints from different probes, in some cases approaching the
cosmic variance limit, can be checked for consistency.
Therefore, the community strongly supports continuing the program of Stage III and Stage IV dark energy experiments, and moving forward as quickly as possible with the construction of LSST and DESI.

\item BUILD A GENERATION IV CMB POLARIZATION EXPERIMENT 

Cosmic microwave background experiments can measure the sum of the neutrino masses and the energy scale of
inflation, as well as constrain exotic physics such as early dark energy and extra neutrino species. After the current
generation of small scale experiments complete data taking near the end of the decade, the community understands that
the next generation experiment -- one that can pin down neutrino masses and the scale of large field
inflationary models -- requires a nationwide coherent effort. Moving from thousands to hundreds of thousands of detector elements will require the involvement of the national laboratories working together with the university community.

\item EXTEND THE REACH  

With small additional investments the dark energy program can be augmented in three important ways. A targeted spectroscopic campaign designed to optimize and calibrate methods of redshift estimation from imaging surveys can enhance their science returns beyond their nominal capabilities~\cite{photoz}. Second, a continued investment in instrumentation R\&D will allow the community to do more science for less money and to be ready 
to pounce on future discoveries. 
Finally, a suite of novel probes of gravity and dark energy can discover signatures of modified gravity and new physics in the dark sector. We have the  opportunity to catalyze the next generation of tests by supporting work at the interface of theory, simulation and data analysis, and making small enhancements to the dark energy survey program.

\end{itemize}

\bibliography{de}

\providecommand{\href}[2]{#2}\begingroup\raggedright\begin{thebibliography}{10}

\bibitem{Carroll:2003wy}
S.~M. Carroll, V.~Duvvuri, M.~Trodden, and M.~S. Turner, ``{Is cosmic speed -
  up due to new gravitational physics?},''
  \href{http://dx.doi.org/10.1103/PhysRevD.70.043528}{{\em Phys.Rev.}
  {\bfseries D70} (2004) 043528},
\href{http://arxiv.org/abs/astro-ph/0306438}{{\ttfamily arXiv:astro-ph/0306438
  [astro-ph]}}.

\bibitem{Chiba:2003ir}
T.~Chiba, ``{1/R gravity and scalar - tensor gravity},''
  \href{http://dx.doi.org/10.1016/j.physletb.2003.09.033}{{\em Phys.Lett.}
  {\bfseries B575} (2003) 1--3},
\href{http://arxiv.org/abs/astro-ph/0307338}{{\ttfamily arXiv:astro-ph/0307338
  [astro-ph]}}.

\bibitem{Hu:2007pj}
W.~Hu and I.~Sawicki, ``{A Parameterized Post-Friedmann Framework for Modified
  Gravity},'' \href{http://dx.doi.org/10.1103/PhysRevD.76.104043}{{\em
  Phys.Rev.} {\bfseries D76} (2007) 104043},
\href{http://arxiv.org/abs/0708.1190}{{\ttfamily arXiv:0708.1190 [astro-ph]}}.

\bibitem{Fierz:1939ix}
M.~Fierz and W.~Pauli, ``{On relativistic wave equations for particles of
  arbitrary spin in an electromagnetic field},''
\href{http://dx.doi.org/10.1098/rspa.1939.0140}{{\em Proc.Roy.Soc.Lond.}
  {\bfseries A173} (1939) 211--232}.

\bibitem{Boulware:1973my}
D.~Boulware and S.~Deser, ``{Can gravitation have a finite range?},''
\href{http://dx.doi.org/10.1103/PhysRevD.6.3368}{{\em Phys.Rev.} {\bfseries D6}
  (1972) 3368--3382}.

\bibitem{Vainshtein:1972sx}
A.~Vainshtein, ``{To the problem of nonvanishing gravitation mass},''
\href{http://dx.doi.org/10.1016/0370-2693(72)90147-5}{{\em Phys.Lett.}
  {\bfseries B39} (1972) 393--394}.

\bibitem{Nicolis:2008in}
A.~Nicolis, R.~Rattazzi, and E.~Trincherini, ``{The Galileon as a local
  modification of gravity},''
  \href{http://dx.doi.org/10.1103/PhysRevD.79.064036}{{\em Phys.Rev.}
  {\bfseries D79} (2009) 064036},
\href{http://arxiv.org/abs/0811.2197}{{\ttfamily arXiv:0811.2197 [hep-th]}}.

\bibitem{Albrecht:2006um}
A.~Albrecht, G.~Bernstein, R.~Cahn, W.~L. Freedman, J.~Hewitt, {\em et~al.},
  ``{Report of the Dark Energy Task Force},''
\href{http://arxiv.org/abs/astro-ph/0609591}{{\ttfamily arXiv:astro-ph/0609591
  [astro-ph]}}.

\bibitem{rocky3}
A.~Albrecht, S.~Dodelson, C.~Hirata, D.~Huterer, B.~Jain, {\em et~al.},
  ``{Community Dark Energy Task Force Report},''
  \href{http://arxiv.org/abs/http://science.energy.gov/hep/hepap/meetings/20120827}{{\ttfamily
  http://science.energy.gov/hep/hepap/meetings/20120827}}.

\bibitem{projects}
D.~Weinberg {\em et~al.}, ``{Facilities for Dark Energy Investigations},''
\href{http://arxiv.org/abs/1309.5380}{{\ttfamily arXiv:1309.5380
  [astro-ph.CO]}}.

\bibitem{novel}
B.~Jain {\em et~al.}, ``{Novel Probes of Gravity and Dark Energy},''
\href{http://arxiv.org/abs/1309.5389}{{\ttfamily arXiv:1309.5389
  [astro-ph.CO]}}.

\bibitem{Zhan:2006gi}
H.~Zhan, ``{Cosmic tomographies: baryon acoustic oscillations and weak
  lensing},'' \href{http://dx.doi.org/10.1088/1475-7516/2006/08/008}{{\em JCAP}
  {\bfseries 0608} (2006) 008},
\href{http://arxiv.org/abs/astro-ph/0605696}{{\ttfamily arXiv:astro-ph/0605696
  [astro-ph]}}.

\bibitem{distance}
A.~Kim {\em et~al.}, ``{Distances},''
\href{http://arxiv.org/abs/1309.5382}{{\ttfamily arXiv:1309.5382
  [astro-ph.CO]}}.

\bibitem{growth}
D.~Huterer {\em et~al.}, ``{Growth of Cosmic Structure: Probing Dark Energy
  Beyond Expansion},''
\href{http://arxiv.org/abs/1309.5385}{{\ttfamily arXiv:1309.5385
  [astro-ph.CO]}}.

\bibitem{cross}
J.~Rhodes {\em et~al.}, ``{Exploiting Cross Correlations and Joint Analyses},''
\href{http://arxiv.org/abs/1309.5388}{{\ttfamily arXiv:1309.5388
  [astro-ph.CO]}}.

\bibitem{DGP}
G.~Dvali, G.~Gabadadze, and M.~Porrati, ``{4-D gravity on a brane in 5-D
  Minkowski space},''
  \href{http://dx.doi.org/10.1016/S0370-2693(00)00669-9}{{\em Phys.Lett.}
  {\bfseries B485} (2000) 208--214},
\href{http://arxiv.org/abs/hep-th/0005016}{{\ttfamily arXiv:hep-th/0005016
  [hep-th]}}.

\bibitem{0905.2962}
E.~V. {Linder}, ``{Exponential gravity},''
  \href{http://dx.doi.org/10.1103/PhysRevD.80.123528}{{\em "Phys. Rev. D}
  {\bfseries 80} no.~12, (Dec., 2009) 123528},
  \href{http://arxiv.org/abs/0905.2962}{{\ttfamily arXiv:0905.2962
  [astro-ph.CO]}}.

\bibitem{Hu:2007nk}
W.~Hu and I.~Sawicki, ``{Models of f(R) Cosmic Acceleration that Evade
  Solar-System Tests},''
  \href{http://dx.doi.org/10.1103/PhysRevD.76.064004}{{\em Phys.Rev.}
  {\bfseries D76} (2007) 064004},
\href{http://arxiv.org/abs/0705.1158}{{\ttfamily arXiv:0705.1158 [astro-ph]}}.

\bibitem{Jain:2012tn}
B.~Jain, V.~Vikram, and J.~Sakstein, ``{Astrophysical Tests of Modified
  Gravity: Constraints from Distance Indicators in the Nearby Universe},''
\href{http://arxiv.org/abs/1204.6044}{{\ttfamily arXiv:1204.6044
  [astro-ph.CO]}}.

\bibitem{Vikram:2013uba}
V.~Vikram, A.~Cabre, B.~Jain, and J.~VanderPlas, ``{Astrophysical Tests of
  Modified Gravity: the Morphology and Kinematics of Dwarf Galaxies},''
\href{http://arxiv.org/abs/1303.0295}{{\ttfamily arXiv:1303.0295
  [astro-ph.CO]}}.

\bibitem{Song:2007da}
Y.-S. Song, H.~Peiris, and W.~Hu, ``{Cosmological Constraints on f(R)
  Acceleration Models},''
  \href{http://dx.doi.org/10.1103/PhysRevD.76.063517}{{\em Phys.Rev.}
  {\bfseries D76} (2007) 063517},
\href{http://arxiv.org/abs/0706.2399}{{\ttfamily arXiv:0706.2399 [astro-ph]}}.

\bibitem{Giannantonio:2009gi}
T.~Giannantonio, M.~Martinelli, A.~Silvestri, and A.~Melchiorri, ``{New
  constraints on parametrised modified gravity from correlations of the CMB
  with large scale structure},''
  \href{http://dx.doi.org/10.1088/1475-7516/2010/04/030}{{\em JCAP} {\bfseries
  1004} (2010) 030},
\href{http://arxiv.org/abs/0909.2045}{{\ttfamily arXiv:0909.2045
  [astro-ph.CO]}}.

\bibitem{Schmidt:2009am}
F.~Schmidt, A.~Vikhlinin, and W.~Hu, ``{Cluster Constraints on f(R) Gravity},''
  \href{http://dx.doi.org/10.1103/PhysRevD.80.083505}{{\em Phys.Rev.}
  {\bfseries D80} (2009) 083505},
\href{http://arxiv.org/abs/0908.2457}{{\ttfamily arXiv:0908.2457
  [astro-ph.CO]}}.

\bibitem{High:2012un}
F.~High, H.~Hoekstra, N.~Leethochawalit, T.~de~Haan, L.~Abramson, {\em et~al.},
  ``{Weak-Lensing Mass Measurements of Five Galaxy Clusters in the South Pole
  Telescope Survey Using Magellan/Megacam},''
  \href{http://dx.doi.org/10.1088/0004-637X/758/1/68}{{\em Astrophys.J.}
  {\bfseries 758} (2012) 68},
\href{http://arxiv.org/abs/1205.3103}{{\ttfamily arXiv:1205.3103
  [astro-ph.CO]}}.

\bibitem{Lyth:1996im}
D.~H. Lyth, ``{What would we learn by detecting a gravitational wave signal in
  the cosmic microwave background anisotropy?},''
  \href{http://dx.doi.org/10.1103/PhysRevLett.78.1861}{{\em Phys.Rev.Lett.}
  {\bfseries 78} (1997) 1861--1863},
\href{http://arxiv.org/abs/hep-ph/9606387}{{\ttfamily arXiv:hep-ph/9606387
  [hep-ph]}}.

\bibitem{inflation}
J.~Carlstrom, A.~Lee, {\em et~al.}, ``Inflation physics from the cosmic
  microwave background and large scale structure,''
\href{http://arxiv.org/abs/1309.5381}{{\ttfamily arXiv:1309.5381
  [astro-ph.CO]}}.

\bibitem{Ade:2013uln}
{\bfseries Planck Collaboration} Collaboration, P.~Ade {\em et~al.}, ``{Planck
  2013 results. XXII. Constraints on inflation},''
\href{http://arxiv.org/abs/1303.5082}{{\ttfamily arXiv:1303.5082
  [astro-ph.CO]}}.

\bibitem{Ade:2013zuv}
{\bfseries Planck Collaboration} Collaboration, P.~Ade {\em et~al.}, ``{Planck
  2013 results. XVI. Cosmological parameters},''
\href{http://arxiv.org/abs/1303.5076}{{\ttfamily arXiv:1303.5076
  [astro-ph.CO]}}.

\bibitem{neutrino}
K.~Abazajian {\em et~al.}, ``{Neutrino Physics from the Cosmic Microwave
  Background and Large Scale Structure},''
\href{http://arxiv.org/abs/1309.5383}{{\ttfamily arXiv:1309.5383
  [astro-ph.CO]}}.

\bibitem{Aguilar:2001ty}
{\bfseries LSND Collaboration} Collaboration, A.~Aguilar-Arevalo {\em et~al.},
  ``{Evidence for neutrino oscillations from the observation of
  anti-neutrino(electron) appearance in a anti-neutrino(muon) beam},''
  \href{http://dx.doi.org/10.1103/PhysRevD.64.112007}{{\em Phys.Rev.}
  {\bfseries D64} (2001) 112007},
\href{http://arxiv.org/abs/hep-ex/0104049}{{\ttfamily arXiv:hep-ex/0104049
  [hep-ex]}}.

\bibitem{AguilarArevalo:2010wv}
{\bfseries MiniBooNE Collaboration} Collaboration, A.~Aguilar-Arevalo {\em
  et~al.}, ``{Event Excess in the MiniBooNE Search for $\bar \nu_\mu
  \rightarrow \bar \nu_e$ Oscillations},''
  \href{http://dx.doi.org/10.1103/PhysRevLett.105.181801}{{\em Phys.Rev.Lett.}
  {\bfseries 105} (2010) 181801},
\href{http://arxiv.org/abs/1007.1150}{{\ttfamily arXiv:1007.1150 [hep-ex]}}.

\bibitem{Kopp:2011qd}
J.~Kopp, M.~Maltoni, and T.~Schwetz, ``{Are there sterile neutrinos at the eV
  scale?},'' \href{http://dx.doi.org/10.1103/PhysRevLett.107.091801}{{\em
  Phys.Rev.Lett.} {\bfseries 107} (2011) 091801},
\href{http://arxiv.org/abs/1103.4570}{{\ttfamily arXiv:1103.4570 [hep-ph]}}.

\bibitem{Giunti:2011gz}
C.~Giunti and M.~Laveder, ``{3+1 and 3+2 Sterile Neutrino Fits},''
  \href{http://dx.doi.org/10.1103/PhysRevD.84.073008}{{\em Phys.Rev.}
  {\bfseries D84} (2011) 073008},
\href{http://arxiv.org/abs/1107.1452}{{\ttfamily arXiv:1107.1452 [hep-ph]}}.

\bibitem{Albrecht:2009ct}
A.~Albrecht, L.~Amendola, G.~Bernstein, D.~Clowe, D.~Eisenstein, {\em et~al.},
  ``{Findings of the Joint Dark Energy Mission Figure of Merit Science Working
  Group},''
\href{http://arxiv.org/abs/0901.0721}{{\ttfamily arXiv:0901.0721
  [astro-ph.IM]}}.

\bibitem{photoz}
J.~Newman {\em et~al.}, ``{Spectroscopic Needs for Imaging Dark Energy
  Experiments},''
\href{http://arxiv.org/abs/1309.5384}{{\ttfamily arXiv:1309.5384
  [astro-ph.CO]}}.

\end{thebibliography}\endgroup



\end{document}